# FingerSlid: Towards Finger-Sliding Continuous Authentication on Smart Devices Via Vibration

Yadong Xie, *Graduate Student Member, IEEE*, Fan Li, *Member, IEEE*, and Yu Wang, *Fellow, IEEE*

*Abstract*—Nowadays, mobile smart devices are widely used in daily life. It is increasingly important to prevent malicious users from accessing private data, thus a secure and convenient authentication method is urgently needed. Compared with common one-off authentication (e.g., password, face recognition, and fingerprint), continuous authentication can provide constant privacy protection. However, most studies are based on behavioral features and vulnerable to spoofing attacks. To solve this problem, we study the unique influence of sliding fingers on active vibration signals, and further propose an authentication system, FingerSlid, which uses vibration motors and accelerometers in mobile devices to sense biometric features of sliding fingers to achieve behavior-independent continuous authentication. First, we design two kinds of active vibration signals and propose a novel signal generation mechanism to improve the anti-attack ability of FingerSlid. Then, we extract different biometric features from the received two kinds of signals, and eliminate the influence of behavioral features in biometric features using a carefully designed Triplet network. Last, user authentication is performed by using the generated behavior-independent biometric features. FingerSlid is evaluated through a large number of experiments under different scenarios, and it achieves an average accuracy of 95.4% and can resist 99.5% of attacks.

*Index Terms*—Continuous authentication, finger sliding, vibration signal.

## I. Introduction

WITH the advent of the mobile Internet era, mobile smart devices are widely used in more and more people's daily life. The amount of personal data stored in these devices (e.g., smartphones and smartwatches) is also increasing. Therefore, the security of these devices become a key concern for a large number of users. According to a report [1], the number of victims involved in data leakage is more than 300 million in 2020. Besides, a survey from Cisco [2] shows that nearly 90% of users are concerned about the security of their private data, and 80% of them would like to take active action to protect it. As a necessary technology to meet the requirements of secure access control of mobile devices, user authentication is widely used in various mobile smart devices.

Currently, the most commonly used mobile smart devices mainly include smartphones and smart wristband devices. For smartphones, authentication using biometric features, including face [3], fingerprint [4], iris [5], voice [6], etc., are widely applied. Different from traditional passwords, biometric authentication is theoretically more secure. Moreover, it relieves the burden on users to remember complicated passwords, but it is vulnerable to replay attacks. And it only works once when users log in to the device, so it is easy for an attacker to steal private data when the user forgets to lock the device. In order to reduce the possibility of a successful attack, continuous authentication is proposed to recheck the user's identity with a high frequency. Therefore, continuous authentication goes a step forward to strengthen system security. Touchalytics [7] is based on the behavioral features of interactions between users and the touch screen to realize continuous authentication. However, the system which is related to behavior is vulnerable to mimic attacks. Crouse et al. [8] combine the Inertial Measurement Unit (IMU) and camera to capture face images and realize a continuous authentication, but it is sensitive to ambient light conditions. At present, the authentication method on COTS wristband devices is mainly passwords, but it is inconvenient to input because of the small screen. Some authentication methods require users to make specific hand movements for authentication, such as waving hands [9], putting up or down arm [10], and drawing names in the air [11]. But these behavior-based methods place high demands on the consistency of user movements. The biometric-based continuous authentication on wristband devices mainly collects heartbeat signals [12], which is susceptible to the physical state. Therefore, a cross-device, behavior-independent, and continuous authentication method is urgently needed to improve the security of mobile smart devices.

When people use mobile smart devices (i.e., smartphones and smartwatches), the most frequent actions are clicking or sliding a finger on the screen. It is known that fingers of different people have unique physical features, such as shape, size, bone density, and muscle distribution [13]. To achieve behavior-independent continuous authentication, we can take advantage of these unique physical features of the sliding finger. In addition, the sensors that exist in most of mobile smart devices are the vibration motor and IMU (e.g., accelerometer). And existing studies [14], [15] show that different human bodies

Manuscript received 4 May 2023; revised 26 August 2023; accepted 11 September 2023. Date of publication 14 September 2023; date of current version 4 April 2024. The work of Fan Li was supported by the National Natural Science Foundation of China (NSFC) under Grant 62072040. Recommended for acceptance by R. Zhang. *(Corresponding author: Fan Li.)*

This work involved human subjects or animals in its research. Approval of all ethical and experimental procedures and protocols was granted by the Ethics Committee of Beijing Institute of Technology under Application No. 2023081, and performed in line with the principles of the Declaration of Helsinki.

Yadong Xie and Fan Li are with the School of Computer Science and Technology, Beijing Institute of Technology, Beijing 100089, China (e-mail: ydxie@bit.edu.cn; fli@bit.edu.cn).

Yu Wang is with the Department of Computer and Information Sciences, Temple University, Philadelphia, PA 19122 USA (e-mail: wangyu@temple.edu).

Digital Object Identifier 10.1109/TMC.2023.3315291

have different effects on the active vibration generated by the vibration motor, resulting in different vibration signals received by the accelerometer. However, the duration of the click action is extremely short (less than 40 ms), so it is difficult to extract enough biometric features of fingers for authentication. On the contrary, the duration of the slide action is usually more than 100 ms, which is suitable for constructing a continuous authen- tication system. Motivated by this, we propose an authentication system, FingerSlid, which uses active vibration signals to sense the biometric features of users' sliding fingers.

However, there are still several practical challenges to realize FingerSlid. First, although we know finger sliding on the screen affects the active vibration signal, it is unclear how such effect can be used for user authentication. Second, in order to improve the anti-attack ability of FingerSlid, we make the vibration signal of each authentication different and unpredictable. Specifically, we add different random frequency segments (RFSs) to the vibration signal of each authentication, then how to accurately locate the frequency and time of each RFS in the received signal becomes a challenge. Third, the behavioral features (e.g., sliding force and path) also affect the vibration signal. We need to design an appropriate method that can eliminate the influence of user behaviors and preserve the inherent physical features of users' fingers. Finally, to meet the user-friendliness, we need to use as little registration data as possible to achieve outstanding system performance.

To address the above challenges, we first study the feasibility of using active vibration signals to capture the unique physical features of sliding fingers. Our preliminary experiments show that sliding fingers have unique effects on two kinds of vibration signals. Then, we design the sending signal as a combination of two different vibration signal components. When detecting the user's finger sliding, FingerSlid generates a sending signal with a variable number of RFSs through the vibration motor, and uses the accelerometer to receive the signal. For the received signal, we first perform signal preprocessing to filter out the interference of device motion and separate two components. To determine whether the RFSs in the received signal are correct, we use the Short Time Energy (STE) and the Fast Fourier Transform (FFT) respectively to check whether the RFSs match in the time and frequency domains. Then, we extract different biometric features from the two signal components. To eliminate the behavioral features, we design an input selection method and a Triplet network [16] to retain only the physical features in the biometric features. Finally, we calculate the center biometric feature (CBF) to identify users accurately.

The prototypes of FingerSlid are built by using linear vibration motors (LRAs), accelerometers, and different types of mobile smart devices. To evaluate the performance, we recruit 40 volunteers (25 males and 15 females) and ask them to use our system in 3 different scenarios. The results demonstrate that FingerSlid can authenticate users accurately in different scenarios, and has

- We study and find that sliding fingers of different individuals have unique effects on active vibration signals because of the discrepant physical features of fingers. We propose a new continuous authentication method, FingerSlid, for mobile smart devices, which uses active vibration signals to sense the biometric features of users' sliding fingers.
- In order to improve the security of FingerSlid, we add several RFSs to each vibration signal. We design a unique algorithm to extract the time and frequency of RFSs from the received signal, so that it can determine whether the RFSs in the sending and received signals match.
- We extract different features from the two vibration signal components, and leverage a Triplet network to extract behavior-independent biometric features to achieve accurate authentication.
- We evaluate FingerSlid via implementing hardware prototypes and conducting extensive experiments in different scenarios. The results show that FingerSlid can identify users with an average accuracy of 95.4%, and defend against various kinds of attacks.

## II. RELATED WORK

We review previous works related to FingerSlid, including biometric-based, vibration-based, and continuous authentications.

*Biometric-Based Authentication:* Biometric-based authentications [23] usually rely on unique physical features inherent to users, such as fingerprint [4], iris [5], face [3], and voice [6]. However, these methods are vulnerable to replay attacks. In addition to these methods, TouchPrint [24] presents an authentication method for smartphones, which relies on the user's hand posture shape traits. SmileAuth [25] extracts dental edge features for user authentication. But this method is sensitive to light. SonicPrint [4] authenticates users by using a microphone to collect the sound of a finger swiping on a smartphone. However, it is easily affected by ambient noise. VoiceGesture [6] detects users by using the articulatory gestures when they are speaking. It transmits a high-frequency sound from the speaker and listens to the reflections at the microphone. But it requires the user to make a voice for authentication, which is not suitable in some quiet environments.

*Vibration-Based Authentication:* Recently, the vibration-based authentication becomes an attractive approach. VELODY [17] uses a vibration speaker to sense the biometrics of a human hand on a vibrating surface. VibWrite [18] is based on a touch-sensing technology, which supports users in using PIN codes, lock patterns, and simple gestures for authentication. Lee et al. [20] and VibID [19] analyze the response to vibrations from a smartwatch, based on the fact that vibrations are propagated differently according to the physical structure of each user. TouchPass [21] uses vibration signals to capture the unique physical features of fingers touching for user authentication. These methods authenticate only once when users unlock the screen, which give attackers an opportunity when the user forgets to lock the device. HandPass [22] employs passive vibration of fingers touching for continuous authentication, but it is affected by the movement of the device. Table I shows the comparison of these methods with our method. These works require users to keep their fingers or hands stationary for a period of time, which makes them unsuitable for continuous authentication.

TABLE I
COMPARISON OF VIBRATION-BASED AUTHENTICATION METHODS

|  | VELODY [17] | VibWrite [18] | VibID [19] | Lee et al. [20] | TouchPass [21] | HandPass [22] | Our Method |
|---|---|---|---|---|---|---|---|
| Device | Solid surface | | Smartwatch | | Smartphone | | Smart device |
| Vibration Frequency | 0.5-10kHz | 16-22kHz | 23-133Hz | 170-240Hz | 130Hz | Null | 150-250Hz |
| Variable Vibration | Yes | No | | | | | Yes |
| Behavior Independent | No | | | | Yes | No | Yes |
| Continuous Authentication | No | | | | Yes | | Yes |
| False Accept Rate | 5.7% | 5% | 3.7% | 1.3% | 2% | 2.3% | **1.3%** |

*Continuous Authentication:* Continuous authentication is the method of confirming users' identity in real-time when they are using mobile smart devices. There are some works that focus on continuous authentication based on behaviors, including keystroke [26], gait [27]. Touchalytics [7] collects geometric patterns as users swipe fingers across the smartphone's screen, including stroke timing, force, and area covered on the screen, to realize continuous authentication. SilentSense [28] uses touch and movement behavior for continuous authentication. However, these works depending on behavioral features are vulnerable to mimic attacks and require users to have very similar behaviors during registration and login. PPGPass [12] leverages PPG sensors in wristband devices to extract biometric features for authentication. But PPG sensors are rarely built in smartphones, they are only deployed on some smart wristband devices.

Different from these works, FingerSlid uses the vibration motor and accelerometer commonly equipped in mobile smart devices to continuously obtain the biometric features of users when their fingers slide on screens. Furthermore, we remove the influence of behavioral features from biometric features.

## III. PRELIMINARY

In this section, we study the propagation model of vibration signal and the feasibility of using active vibration signal for authentication. Finally, we exposit the attack models.

### A. Propagation Model of Active Vibration Signal

As the main medium for human contact with external objects, fingers of people have different physical features, such as shapes, sizes, and bone densities [13]. After research, we find two physical models that can analyze the impact of sliding finger on the propagation of vibration signals.

*1) Mass-Spring-Damper Model:* We first introduce the Mass-Spring-Damper (MSD) model, which is well-suited for modeling objects with complex material properties. Mass, stiffness, and damping are the most important dynamic properties of a mechanical system [29]. Since vibration is affected by all these properties, perfect modeling of vibration is a very complex problem. To simplify the problem, we assume a single-degree-of-freedom model for a sliding finger, which can be easily analyzed. The model describes that displacement $x(t)$ under a changeable external force $f(t)$ satisfies:

$$m\frac{d^2x(t)}{dt^2} + b\frac{dx(t)}{dt} + kx(t) = f(t), \quad (1)$$

where $m$, $b$, and $k$ are mass, damper coefficient, and spring coefficient of a finger, respectively. Further, the damping ratio $\xi$ is computed as $\xi = b/(2\sqrt{k \cdot m})$.

If the external force is $f(t) = Ae^{i\omega t}$, where $A$ and $\omega$ are the amplitude and frequency, the solution of (1) is:

$$x_f(t) = \frac{A}{k}\frac{1}{\sqrt{\left(1-\frac{\omega^2}{\omega_n^2}\right)^2 + \left(\frac{2\xi\omega}{\omega_n}\right)^2}}e^{i(\omega t+\theta)}, \quad (2)$$

where $\omega_n = \sqrt{k/m}$ is the natural frequency of the finger, and $\theta$ is the phase delay. Different person's fingers have different physical features, resulting in unique $\xi$ and $\omega_n$ [13], so there are different $m$, $b$, and $k$, which ultimately have a significant influence on the vibration response.

*2) Non-Linear Model:* Besides being a complex mechanical system, the human hands and fingers are also non-linear mediums for vibration propagation [30]. When a vibration signal with a constant frequency $f$ is applied to a linear medium, the vibration response only contains the frequency $f$. However, once the medium possesses non-linearity, the vibration response contains not only the frequency component $f$, but also its harmonics, whose frequencies are positive integers multiple of the frequency $f$. Specifically, we can model the non-linear response $S(t)$ as a power-series of vibration signal $f(t) = \sin(2\pi ft)$ with different gains $A_k$:

$$S(t) = \sum_{k=1}^{N} A_k [\sin(2\pi ft)]^k, \quad (3)$$

where $N$ is the order of harmonics.

The harmonics are sensitive to the properties of the vibration medium (i.e., fingers), thus it is very hard to predict the detailed composition of harmonics due to complicated energy exchange between different mediums and temporal dependencies of non-linear coefficients [31].

### B. Feasibility Study

According to the above analysis, we can know that the different fingers have different effects on the propagation of active vibration signals because of their unique physical features. Since it is very complex to analyze the vibration model composed of fingers and mobile smart devices, we conduct experiments to verify the feasibility. We use a vibration motor to generate vibration signals and an accelerometer to collect vibration responses. We ask 3 volunteers to use fingers to slide on the screen of a smartphone. To eliminate the influence of user behaviors, we require each user to keep the sliding force, sliding duration, and sliding path as consistent as possible.

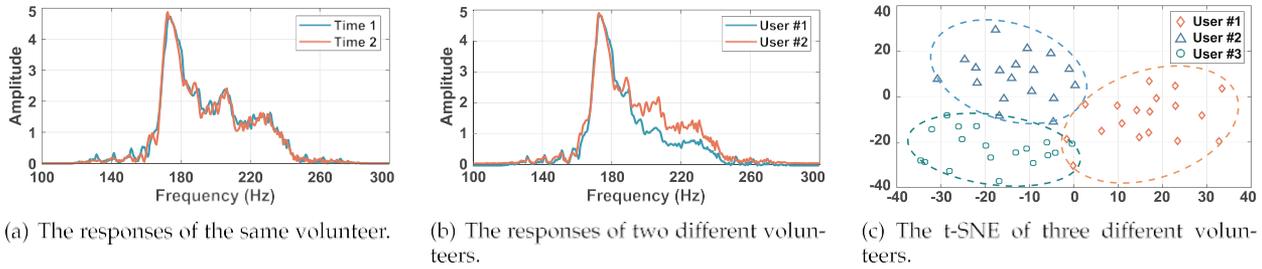

Fig. 1. Feasibility analysis of the chirp vibration signal.

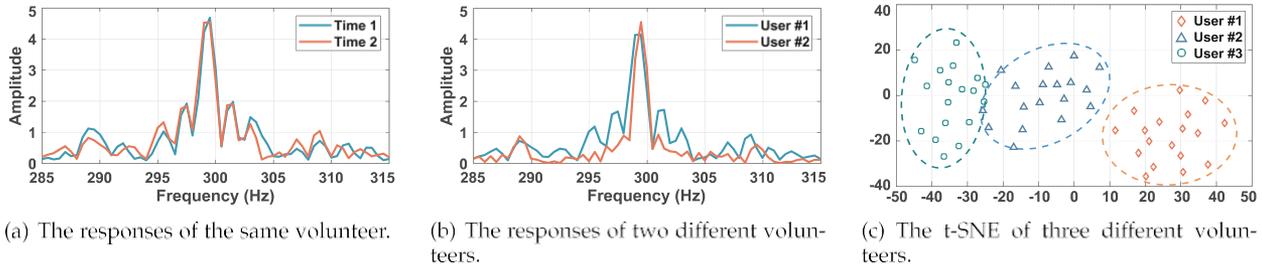

Fig. 2. Feasibility analysis of the stable vibration signal.

*1) Feasibility of Chirp Signal for MSD Model:* According to the theoretical analysis in Section III-A1, the vibration response of MSD model is mainly affected by human damping rate and natural frequency, and the vibration response is more obvious when the vibration frequency is equal to the natural frequency. But the natural frequencies of different human bodies vary greatly [31] because of their unique physical features. To cover the natural frequency range of most users, we use a chirp signal for experiments, the frequency of which increases with time, and the frequency range is from 150 Hz to 250 Hz. After collecting the data, we process each vibration response by using FFT. From Fig. 1(a), we can observe that responses from the same volunteer have similar patterns in the frequency domain. Fig. 1(b) shows that the responses from different volunteers have significant differences. Then, we use the t-SNE method to map the responses of three volunteers into a two-dimensional space [32]. As shown in Fig. 1(c), most of the responses from the same volunteer are concentrated in a nearby area and separated from other volunteers. In addition, the vibration responses of different volunteers overlap slightly, so we need to extract more effective features (detailed in Section IV-E1) to maximize the differences between users.

*2) Feasibility of Stable Signal for Non-Linear Model:* Different human fingers, as non-linear media for the propagation of vibration signals, have different effects on the harmonics of the vibration signals. To extract the influence of the finger's non-linear properties on vibration signals, we send a vibration signal with a constant frequency of 150 Hz. After calculating the FFT, we find that at the second harmonic (i.e., 300 Hz), the vibration responses of different volunteers have significant differences. As shown in Fig. 2(a) and (b), the vibration responses of the same volunteer are very similar at the harmonic, and the two volunteers have clearly distinguishable patterns in the frequency domain. We also use t-SNE to visualize the vibration responses of three volunteers to stable signals, as shown in Fig. 2(c). It can be seen that the vibration responses of the three volunteers can be well distinguished.

Through the feasibility experiments, we come to the conclusion that different person's fingers can produce significantly distinguishable responses to different forms of active vibration signals (i.e., chirp signal and stable signal), which can be used as the basis of FingerSlid.

### C. Attack Model

Nowadays, user authentication systems are usually vulnerable to replay attacks and mimic attacks. Furthermore, we consider an extreme scenario in which the attacker has authentication information and device permissions of the legitimate user, so we design the advanced attack.

*Replay Attack:* In a replay attack, the attacker can attack the device by surreptitiously recording authentication information used by legitimate users. For our system, the attacker secretly places an accelerometer next to the mobile smart device of a legitimate user to record the vibration signal of the authentication process. Then he/she uses a vibration motor to replay the recorded signal to our system for an attack.

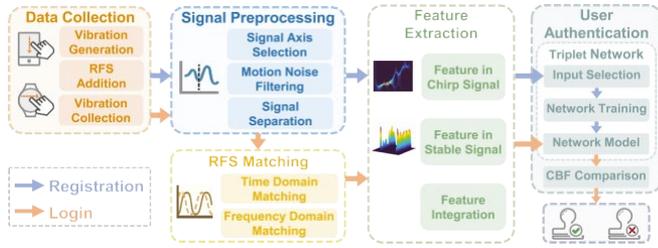

Fig. 3. System architecture of FingerSlid.

*Mimic Attack:* In a mimic attack, the attacker obverses the authentication process of legitimate users and tries his/her best to imitate the finger-sliding behavior of the legitimate user. The attacker needs to control the force, path, and duration of the finger sliding motion, which are basically the same as that of a legitimate user, to deceive our system.

*Advanced Attack:* In an advanced attack, we assume attackers have permissions for the vibration motor and the IMU in a legitimate user's device, and know that our system has an RFS mechanism (detailed in Section IV-B). They can collect the received signal of each authentication through the IMU, and find the start time and frequency of RFSs in each received signal, then filter the RFSs in the received signal. After filtering, the received signal only contains the legitimate user's biometric features. However, the RFSs generated in each authentication process are different, so the attackers should predict the RFSs of the next sending signal by mathematical modeling (e.g., linear regression), and add predicted RFSs to the received signal. Finally, they use the newly generated signal played by the vibration motor in a legitimate user's device to attack our system.

## IV. SYSTEM DESIGN

In this section, we introduce the overview of FingerSlid and describe the key techniques for each part in detail.

### A. System Overview

Fig. 3 shows the architecture of FingerSlid. The whole system mainly includes 5 parts.

In the registration phase, we first design an active vibration signal, which contains a chirp signal segment and a stable signal segment. After the device detects a sliding finger, the vibration motor generates a vibration signal, and the unique vibration response affected by the user is received by the accelerometer. FingerSlid executes *Signal Preprocessing* to preprocess the received vibration signal. Specifically, we use the signal-to-noise ratio (SNR) to determine the signal axis for subsequent processing. Then we apply a band-pass filter to remove the noise caused by the movement of the device. We calculate the frequency band variance to separate two signal segments. When it comes to *Feature Extraction*, we use Synchrosqueezed Wavelet Transform (SWT) and Empirical Wavelet Transform (EWT) to extract the unique biometric features of users embedded in two signal segments. Then, we train a Triplet network for user authentication. Specifically, we first construct the input for training using the extracted features. After training, the Triplet network can output behavior-independent biometric feature. Finally, we generate a center biometric feature (CBF) for each registered user by calculating the average of the biometric features. The network model and the CBF are stored in the database for the login phase.

After registration, we authenticate each time the finger slides on the screen, which is called login. It is similar to registration in most of the process. To prevent attacks in the login phase, we design the RFS mechanism. We add RFSs to the sending signal in *Data Collection*, then we determine whether the RFSs in the received signal match the sending signal in *RFS Matching*. Specifically, we analyze the STE to detect whether the RFS appears at the right time. Then we use FFT to get the spectrum of the received signal to detect whether the frequency of each RFS is correct. After extracting the biometric features, we input them into the trained Triplet network to obtain the behavior-independent biometric feature, and compare it with CBFs stored in the system to achieve authentication.

### B. Data Collection

It consists of *Vibration generation and collection* and *RFS addition*, and only the login phase executes *RFS addition*.

*1) Vibration Generation and Collection:* In order to extract the finger's biometric features for authentication, we first need to design the active vibration signal. Nowadays, the types of vibration motors in most mobile smart devices are Linear Resonant Actuators (LRA), which can be adjusted in frequency and amplitude. Fingers have different effects on different types of vibration signals, so we combine two types of vibration signals (i.e., chirp signal and stable signal) to obtain more biometric features of fingers.

Once the device senses a sliding finger on the screen, the LRA first generates a chirp signal [33]. Typically for LRAs, resonant frequencies are around 175 Hz-235 Hz [34]. And when the LRA is driven near the resonant frequency, it can vibrate with a perceptible force. Thus, we set the frequency range of the chirp signal to 150 Hz-250 Hz. Through observing finger sliding motions of people, we find that most sliding durations are greater than 180 ms. To ensure that the chirp signal is sent completely, we set the duration of the chirp signal to 150 ms. After sending the chirp signal, the LRA generates a stable signal to capture the non-linear biometric features of fingers. The stable signal is a cosine wave that contains only a single frequency of 150 Hz. The LRA continues to generate the stable signal until the user's finger stops sliding, so the stable signal has no fixed duration. The complete vibration signal

$$s(t) = \begin{cases} \cos\left(2\pi \frac{f_e - f_s}{2t}t^2 + f_s t\right), & t < 150 \text{ ms}, \\ \cos(2\pi f_c t), & t > 150 \text{ ms}, \end{cases} \quad (4)$$

where $f_s$ and $f_e$ represent the lowest and highest frequency limit of the chirp signal, respectively. And $f_c$ denotes the frequency of the stable signal.

We then use the accelerometer in the same device to receive the vibration responses. The sampling rate is set to 1000 Hz,

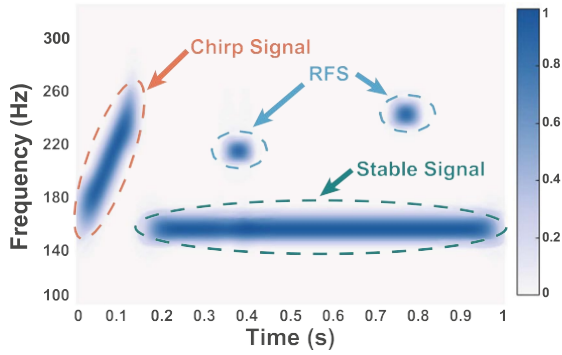

Fig. 4. Spectrum of a complete sending signal.

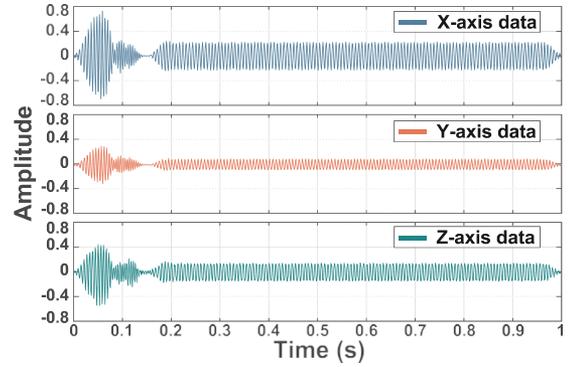

Fig. 5. Vibration responses on 3 axes of the accelerometer.

which does not exceed the upper limit of the sampling rate supported by mobile smart devices.

*2) RFS Addition:* The above-mentioned vibration signal is mainly designed to capture detailed biometric features of fingers. This signal can theoretically resist most replay and mimic attacks. To further improve the anti-attack capability of our system, we design an RFS mechanism by adding RFSs to the vibration signal during the login phase. The RFS is a cosine wave which superimposes onto the stable signal segment. In order not to affect the biometric features contained in the stable signal segment and its harmonics, we set a short duration of 30 ms for each RFS, the frequency range of each RFS is $[100, 135] \cup [165, 250] Hz$, and the number of RFSs is no more than 3 in each authentication. The interval between any two RFSs is greater than 15 ms to ensure distinguishability. Fig. 4 shows the spectrum of a complete vibration signal generated by the LRA in the login phase, which contains three parts: a chirp signal, a stable signal, and two RFSs.

By using the RFS mechanism, each sending signal contains a variable number of RFSs. Then FingerSlid determines whether the vibration response is true by matching the RFSs in the vibration response and the RFSs in the sending signal (detailed in Section IV-D). Even if the attacker can steal the previous vibration responses, he/she can not predict the RFSs correctly in the next sending signal and thus can not use the previous vibration responses to spoof FingerSlid.

### C. Signal Preprocessing

After the user's finger stops sliding, we need to preprocess the collected vibration responses.

*1) Signal Axis Selection:* Since the vibration response collected by the accelerometer contains data in three axes, we need to select the axis that is most sensitive to the vibration signal. For instance, Fig. 5 shows the vibration responses of the three axes collected by the accelerometer. We can see that for each acceleration axis, the collected vibration response is different in amplitude. Since different types of LRAs can vibrate at different axes (i.e., X-axis and Z-axis), and although the LRA and accelerometer are integrated into smart devices at the factory, we can not know their relative position in each model of device. Therefore, FingerSlid needs to automatically select the optimal axis of vibration response.

We leverage SNR to measure the sensitivity of each accelerometer axis to the vibration signal. SNR is defined as the ratio of vibration response power to the noise power. By calculating the SNR of the vibration response on each axis, we can get the most sensitive axis $A_i$:

$$A_i = \arg\max_{i \in \{x,y,z\}} \lg \frac{P_s^i}{P_n^i}, \qquad (5)$$

where $P_s^i$ and $P_n^i$ are the power of the vibration response and noise signal on $i$ axis, respectively. Since the accelerometer continuously collects data for applications such as step counting, we regard the acceleration data within 1 s before the user slides as the noise signal. After that, the vibration response of the most sensitive axis is selected for further processing.

*2) Motion Noise Filtering:* The vibration response is not only affected by the sliding finger, but also by the motion of the user. For instance, when a user uses a smart device while walking, the accelerometer can also record the walking-induced vibration noise. Thus, we need to filter motion noise in the vibration response to minimize their influence. After analyzing the acceleration data in users' daily life, we find that most of the vibration frequency generated by daily activities is less than 80 Hz. And the biometric features of sliding fingers are mainly extracted from frequencies above 150 Hz. So we adopt a high-pass filter with a cutoff frequency of 100 Hz to filter out the interference of low-frequency motion noise.

*3) Signal Separation:* Since we use two kinds of vibration signals (i.e., chirp signal and stable signal) to capture different biometric features of sliding fingers, we design two kinds of feature extraction algorithms (detailed in Section IV-E). Thus, we need to determine the separation point between the two kinds of signals. The straightforward method is to separate the received signal at 150 ms. However, it is inaccurate since the LRA and the accelerometer can not start working perfectly at the same time. The commonly used synchronization method for sending and received signals is to add a short vibration impulse at the beginning of sending signal [18], but it reduces the time for the vibration signal to collect biometric features.

We find that the energy of the chirp signal varies greatly with frequency, while the energy of the stable signal is only concentrated in a small frequency band. Thus, we design a separation method based on frequency band variance. Specifically, we apply a hamming window with a length of 10 ms that slides 2 ms each time on the signal. The vibration signal of the $k$-th window is $x_k(n)$. Its amplitude $|X_k(i)|$ obtained after FFT is evenly divided into $r$ bands, and each band contains $s$ frequency points. The band can be expressed by

$$F_k(m) = \sum_{i=1+(m-1)s}^{1+(m-1)s+(s-1)} |X_k(i)|, m \in [1, r], \quad (6)$$

and its variance can be expressed by

$$D_k = \frac{1}{r-1} \sum_{i=1}^{r} \left( F_k(i) - \frac{1}{r} \sum_{i=1}^{r} F_k(i) \right)^2. \quad (7)$$

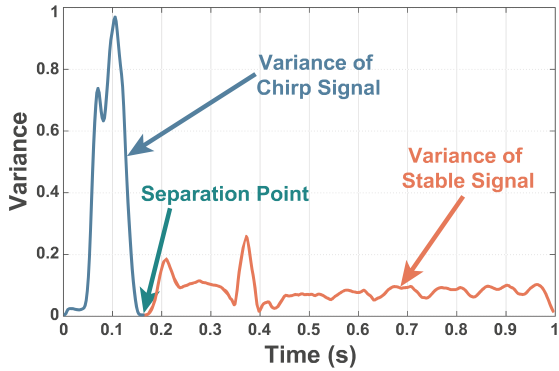

Fig. 6. Frequency band variance of a received signal.

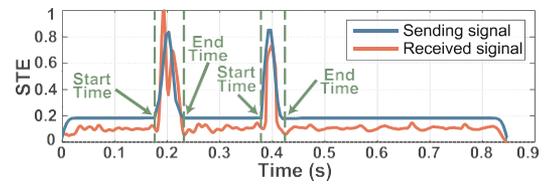

(a) RFS matching in the time domain.

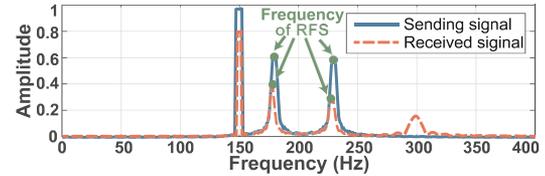

(b) RFS matching in the frequency domain.

Fig. 7. RFS matching in time and frequency domain.

We can observe from (7) that the greater the frequency fluctuation, the greater the $D_k$. Fig. 6 shows the frequency band variance of a received signal, there is a minimum value of variance between the chirp signal and the stable signal. Thus, we can determine the location of the separation point by searching for the minimum point of the variance.

### D. RFS Matching

During the login phase, RFSs are added to the stable signal segment to prevent attacks. After generating sending signal, FingerSlid can obtain the specific time-frequency information of RFSs. Then, the accelerometer receives the sending signal, FingerSlid needs to determine whether the RFSs in the received signal match the RFSs in the sending signal. If they have the same time-frequency information, it means that the received signal is indeed generated by this authentication.

*1) Time Domain Matching:* We first perform RFS matching in the time domain, that is, determine the start and end time of each RFS. After observing the received signal, we find that the signal energy is significantly higher with the presence of RFSs than without RFSs. Thus, we can use STE [35], which is usually used to distinguish between voiced and unvoiced segments in speech analysis, to determine the start and end time of each RFS. Specifically, we first perform windowing and framing on the stable signal $y(n)$, then we can get the $k$-th frame signal $y_k(n)$ and its STE as $E_k = \sum_{n=0}^{l-1} y_k^2(n)$, the length of each frame $l$ is 10 ms.

After calculating the STE, we design a method based on the energy difference of STE to detect the start and end time of each RFS. First, we calculate the energy difference between two frames according to $D_t = E_{t+1} - E_t$. Then, we set two thresholds of the energy difference for the start time and end time, which are $\eta_s$ and $\eta_e$ respectively. When $D_t > \eta_s$ and $E_{t+2} > E_{t+1}$, we take $t$ as the start time of the candidate RFS, and then look for the corresponding end time. When $D_t < \eta_e$ and $E_t < E_{t-1}$, we take $t$ as the end time of the candidate RFS. Then we look for the start time of the next candidate RFS until we find all the start and end times.

However, some sudden vibrations may affect the STE and be detected. But the STE of these vibrations is usually less than that of RFS, so we find the energy peaks of candidate RFSs and select the top $k$ RFSs with the largest energy ($k$ is the number of RFSs in the sending signal). Finally, we calculate the time difference between the start/end time of the selected RFS and the RFS in the sending signal. If the time difference is less than 5 ms, we consider that the signals match in the time domain. Fig. 7(a) shows the STE of sending and received signals, it can be seen that the RFSs of the two signals are matched.

*2) Frequency Domain Matching:* After the RFS matching in the time domain, we match the frequencies of RFSs in the sending and received signals. Since RFSs are only added to the stable signal segment, we use FFT to analyze the frequency distribution of the signal. Fig. 7(b) shows the spectrum of the stable signal with RFSs, we can find that the highest amplitude on the spectrum is the frequency of the stable signal (i.e., 150 Hz). Since the RFS has a fixed frequency and relatively high energy, it also shows a relatively high peak in the spectrum. We use the peak detection algorithm to get all peaks in the spectrum and sort them. Then we find the top $k$ peaks with the largest amplitudes except for the peak at the 150 Hz, so we can get the corresponding frequencies of RFSs. Finally, we calculate the frequency difference between the selected RFS and the RFS in the sending signal. If the frequency difference is less than 5 Hz, we consider that the signals match in the frequency domain. We can see from Fig. 7(b) that two relatively high peaks correspond to two matching RFSs frequencies in the sending and received signals.

*E. Feature Extraction*

In order to obtain unique biometric features of different users from received signals, we design two feature extraction methods for different segments of the received signals (i.e., chirp signal and stable signal).

*1) Feature Extraction on Chirp Signal:* First, we need to extract biometric features from the chirp signal. In Section III-B1, we prove that chirp signals can well capture the physical features of users and present them in the frequency domain. However, sliding is a dynamic process, so time-domain information of the signal also contains biometric features. At present, the most commonly used analytical methods are Short-time Fourier transform (STFT) [36] and continuous wavelet transform (CWT) [37]. Due to the fixed length of the window, it is difficult for STFT to satisfy both high resolutions in time and frequency. Although CWT can adjust time resolution adaptively, its time-frequency resolution is still not enough to extract fine-grained biometric features from the very short chirp signal (only 150 ms).

To capture the biometric features contained in the chirp signal, we study SWT [38] which has more explicit features in the time-frequency domain. SWT reallocates the signal energy only in the frequency direction, which preserves the time resolution of the signal and compensates for the spreading effects caused by the mother wavelet. For an input signal $s(t)$, we first calculate its CWT $W(a, b)$, where $a$ is the scale factor, and $b$ is the time translation. Then we use the phase transformation to extract the instantaneous frequency $\Omega(a, b)$ of each point on the time-scale plane, which is defined as

$$\Omega(a, b) = -i W(a, b)^{-1} \frac{\partial W(a, b)}{\partial b}, \quad (8)$$

where i is the imaginary unit. After obtaining $\Omega(a, b)$, SWT transfers the information from the time-scale plane to the time-frequency plane. Considering $\Omega_l$ as the closest frequency to $\Omega(a, b)$, each value of $W(a, b)$ is reallocated into $T(\Omega_l, b)$ as

$$T(\Omega_l, b) = (\Delta\Omega)^{-1} \sum_{a_k: |\Omega(a_k, b) - \Omega_l| < \Delta\Omega/2} \times W(a_k, b) a_k^{-3/2} (\Delta a)_k, \quad (9)$$

where $\Delta\Omega = \Omega_l - \Omega_{l-1}$ and $\Omega_l$ is the discrete frequency variable. $(\Delta a)_k = a_k - a_{k-1}$ and $a_k$ is the discrete wavelet scale. Fig. 8 shows two users' biometric features on the chirp signal, which can be clearly distinguished.

*2) Feature Extraction on Stable Signal:* Then we extract biometric features on the stable signal. The feasibility study shows that the biometric features contained in the stable signal are mainly stored in the frequency around 150 Hz and 300 Hz. The Hilbert-Huang transform based on EMD [39] can adaptively conduct time-frequency analysis, so as to extract biometric features. However, the EMD method lacks a complete theoretical foundation and has the problem of modal aliasing. So we explore another time-frequency analysis method, EWT [40], which combines the advantages of EMD (e.g., adaptability and good time-frequency focusing) and the theoretical framework of wavelet analysis.

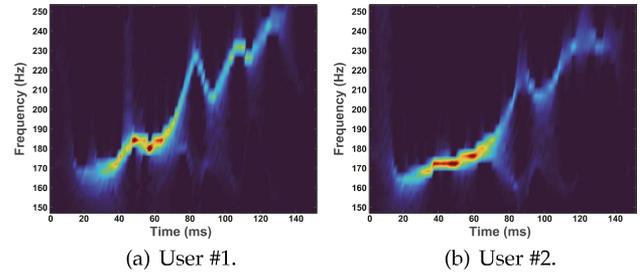

Fig. 8. Feature extraction on chirp signal.

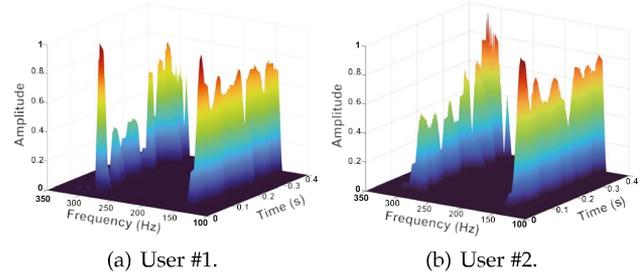

Fig. 9. Feature extraction on stable signal.

The basic idea of EWT is to adaptively select the appropriate orthogonal wavelet filter bank to extract signal modes according to the spectral characteristics of the signal, and then perform the Hilbert transform to obtain the instantaneous frequency and amplitude. Based on the predefined empirical scaling function $\varphi_n(t)$ and the empirical wavelets $\psi_n(t)$, a signal $f(t)$ can be adaptively decomposed as

$$f(t) = W_f(0, t) * \varphi_1(t) + \sum_{n=1}^{N} W_f(n, t) * \psi_n(t), \quad (10)$$

where $W_f(0, t)$ is the approximation coefficient and $W_f(n, t)$ is the detail coefficient. They are obtained by:

$$W_f(0, t) = \langle f, \varphi_1 \rangle = \int f(\tau) \overline{\varphi}_1(\tau - t) d\tau, \quad (11)$$

$$W_f(n, t) = \langle f, \psi_n \rangle = \int f(\tau) \overline{\psi}_n(\tau - t) d\tau. \quad (12)$$

After obtaining each mode $f_k(t)$ of the signal which can be expressed by $f_k(t) = W_f(k, t) * \psi_k(t)$, we perform Hilbert transform on each mode. It extracts the instantaneous frequency and amplitude that can characterize the biometric features. Fig. 9 shows two users' biometric features extracted by EWT in the frequency around 150 Hz and 300 Hz, and we can see that there are obvious differences.

After extracting the biometric features in the chirp signal and stable signal, we need to integrate them together. We splice the two feature matrices together to form a complete feature matrix, and input the feature matrix into the subsequent network model for training and authentication.

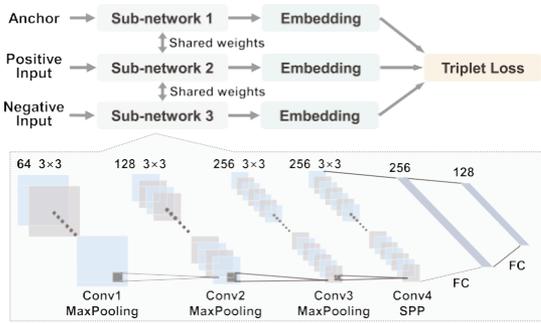

Fig. 10. Structure of the Triplet network.

### F. User Authentication

The feature matrix contains the user's behavioral features and physical features. There are three behavioral features of sliding fingers: sliding duration, force, and path. To achieve behavior-independent authentication, we need to eliminate the influence of behavioral features. Traditional classifiers (e.g., SVM, RF, and CNN) usually have a fixed number of output classes and need a large number of positive and negative data for training. So they are not very suitable for solving the problem that the number of output classes is uncertain and the training data set is small, such as user authentication. Inspired by face recognition [41], we study a novel network, Triplet network, to reconstruct biometric features, making it contain only the physical features inherent to the user's finger. The Triplet network consists of three identical sub-networks, whose input is a triplet consisting of an anchor, a positive input, and a negative input, as shown in Fig. 10. For example, the feature matrix of user $U_1$ under behavior $B_1$ is an anchor, the feature matrix of $U_1$ under $B_2$ is a positive input, and the feature matrix of $U_2$ under $B_1$ is a negative input. The output of each sub-network is an embedding, that is, the behavior-independent biometric feature. The training goal is to minimize the distance between two embeddings of the anchor and positive input, and maximize the distance between two embeddings of the anchor and negative input.

*1) Model Generation: Input Selection:* Before training the Triplet network, we need to construct the inputted triplet. To determine whether the two sliding motions have the same behavior, we first obtain the duration, average force, and start-end positions of each sliding motion from the touch screen sensor. Then, we calculate the difference between the durations of two sliding motions. If the difference is greater than a threshold (obtained by empirical study), it is considered that the two motions contain different behaviors. We also implement the same mechanism for the average force and start-end positions. Finally, we construct the inputted triplet for training the network.
*Network Training:* Fig. 10 also shows the structure of the three identical sub-networks. Each sub-network consists of 4 convolution (Conv) layers, 3 max-pooling layers, 1 Spatial Pyramid Pooling (SPP) layer, and 2 fully connected (FC) layers. The Conv layer is used to extract behavior-independent biometric features, and the max-pooling layer is used to compress the features to simplify the computational complexity. We add an SPP layer between the Conv layer and the FC layer to ensure that the input of the FC layer has a fixed dimension. To prevent overfitting, we add a Batch Normalization (BN) layer after each max-pooling layer. The triplet loss of the network can be expressed by

$$L = \max(d(a, p) - d(a, n) + \text{margin}, 0), \quad (13)$$

where $d(a, p)$ represents the distance between two reconstructed feature vectors from sub-network 1 and sub-network 2, $d(a, n)$ represents the distance between two reconstructed feature vectors from sub-network 1 and sub-network 3. By minimizing the triplet loss, we can achieve the extraction of behavior-independent biometric features.

Before the network is deployed, we pre-train it to handle the single-user situation. We ask 3 volunteers to slide their fingers on different areas of the smart devices' screens with different forces. After collecting volunteers' data and extracting features, we construct triplets and input them into the Triplet network for pre-training. The network calculates triplet loss according to (13), and continuously adjusts network parameters. Then, the pre-trained Triplet network can extract behavior-independent biometric features of different users.

*2) Registration and Login:* In the registration phase, a new user needs to slide several times. The features of the new user, 3 volunteers, and other registered users (if any) are used to form triplets, which are sent to the network for retraining it. Based on incremental learning, we only use the triplets to fine-tune the network parameters, rather than retrain all parameters. FingerSlid then generates a unique CBF for the user, which is the average of behavior-independent biometric feature vectors. Note that the three sub-networks are completely consistent. After registration, we only use one trained sub-network to obtain the behavior-independent feature vector in the login phase.

*Login includes two stages:* the authentication when unlocking the device and the continuous authentication during each interaction. Assuming that there are $x$ registered users and corresponding $x$ CBFs. When a user unlocks the device, we calculate $x$ distances between the feature vectors extracted from the user and all CBFs. Then the current user is identified as the user with the CBF corresponding to the minimum distance. If the minimum distance is larger than a threshold, the current user is considered as an attacker. At each interaction, we calculate the feature of the current user and compare it with the CBF of the user determined at the unlocking. If the distance is still less than the threshold, the current user can continue to use the device, otherwise return to the unlock interface. The location of the device and the information of the last operation are sent to the user's secure mailbox.

## V. IMPLEMENTATION AND EVALUATION

In this section, we introduce the implementation of FingerSlid and show its performance in detail.

### A. Experiment Setup

We implement FingerSlid on different devices (including a smartphone and a smartwatch), as shown in Fig. 11. But due to hardware limitations, we can not fully control the embedded

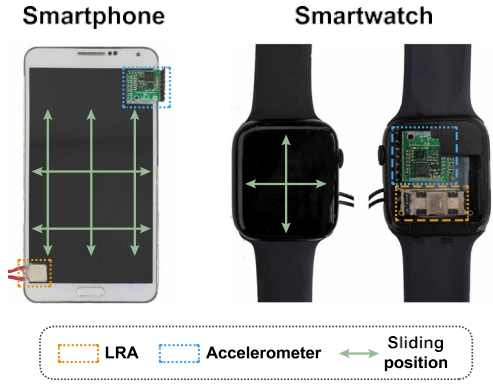

Fig. 11. Prototypes of FingerSlid.

LRA, so we use the extra LRAs and accelerometers fixed to the devices. The LRAs are the same as those built in Xiaomi 9 Pro and Apple Watch S4. Experiments are conducted under 3 scenarios with different levels of motion noise. We recruit 40 volunteers, 30 (18 males and 12 females) of whom are legitimate users and the other 10 are attackers. Legitimate users are asked to slide fingers at 7 start-end positions as shown in Fig. 11 with 3 sliding forces (including light, moderate, and hard). Note that horizontal sliding at the top of the smartphone rarely occurs, so it is not considered. In addition, we consider 2 supports (i.e., desktop and hand) when FingerSlid is deployed on the smartphone. During the 4-month experiments, we collect over 21,000 times sliding for training and testing. Since FingerSlid is a behavior-independent authentication system, we use the data under part of behaviors as the registration data and the data under other behaviors as the test data. Attackers perform 3 kinds of attacks against each legitimate user, and the dataset of each type of attack is no less than 500. All procedures are approved by the Institutional Review Board (IRB) at our institute.

### B. Evaluation Methodology

To evaluate FingerSlid, we consider the following metrics:

*Confusion Matrix:* Each row and each column of the matrix represent the ground truth and the authentication result, respectively.

*F1-Score:* F1-score is defined as the harmonic mean of precision and recall, and its expression is $F_1 = 2 \cdot \frac{precision \cdot recall}{precision+recall}$.

*False Reject Rate (FRR):* The probability that FingerSlid authenticates a legitimate user as a attacker.

*False Accept Rate (FAR):* The probability that FingerSlid authenticates a attacker as a legitimate user.

### C. Overall Performance

We first evaluate the overall performance of FingerSlid for 30 legitimate users (denoted as $U_1, U_2,..., U_{30}$) and 10 attackers (denoted as $AT$) on two prototypes. The confusion matrix in Figs. 12 and 13 show that the average authentication accuracy of FingerSlid for legitimate users on the smartphone and the smartwatch are 95.9% and 93.1%, respectively. Besides, the average accuracy for attacking detection are 99.6% and 99.4%, respectively. Experiment results show that FingerSlid can accu- rately authenticate legitimate users and detect attackers on both smartphone and smartwatch. Compared with the smartwatch, FingerSlid achieves higher accuracy on the smartphone, which may be due to the difference in screen size. Users usually slide on the smartphone for a longer time than they slide on the smartwatch, so that the system can obtain more users' biometric features and further improve the authentication accuracy.

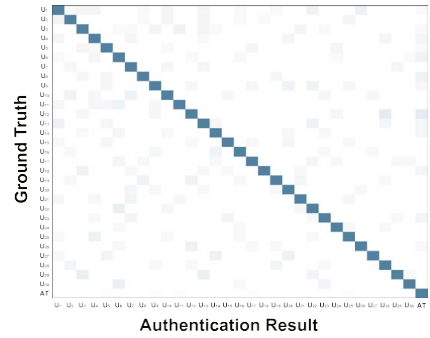

Fig. 12. Confusion matrix on the smartphone.

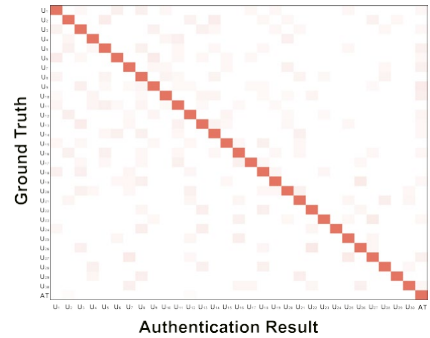

Fig. 13. Confusion matrix on the smartwatch.

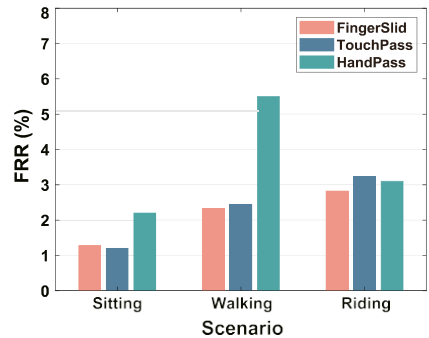

Fig. 14. F1-score under different scenarios.

Then we evaluate FingerSlid with two comparison methods (i.e, TouchPass [21] and HandPass [22]) under three scenarios (i.e., sitting in office, walking outside, and riding in cars). TouchPass uses active vibration to capture features of fingers touching for one-off authentication. HandPass employs passive vibration of hand activated by the fingers touching for continuous authentication. Fig. 14 shows the FRR of three methods in three scenarios. When the user is sitting, the FRR of

TABLE II
F1-SCORE UNDER DIFFERENT SUPPORTS AND TIGHTNESS

| Support/Tightness | Smartphone | | Smartwatch | | |
|---|---|---|---|---|---|
| | Table | Hand | Loose | Fit | Tight |
| F1-score | 95.4% | 94.6% | 92.8% | 93.1% | 90.8% |

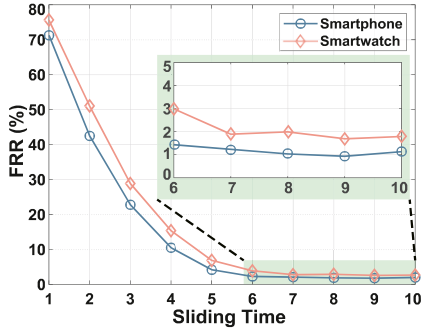

Fig. 15. FRR of sliding times for registration.

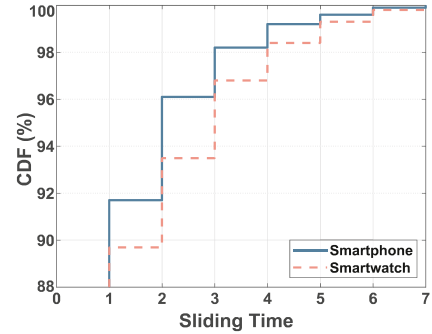

Fig. 16. CDF of sliding times for successful login.

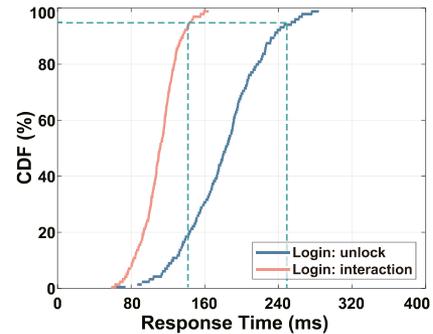

Fig. 17. CDF of response time for login.

In the other scenarios, however, the effect of FingerSlid is better than the other two methods. Overall, FingerSlid performs slightly better than TouchPass. And TouchPass requires the user's finger to remain stationary for 300 ms, so it is not suitable as continuous authentication method. Although HandPass realizes continuous authentication, due to the use of passive vibration, it is obviously affected by the movement of the device. Thanks to the noise reduction algorithm, FingerSlid has good performance in various scenarios.

In addition, users may have different supports when using smartphones, including hand and table. The support of smartwatches is often the wrist, but the tightness of the strap, including loose, fit, and tight, may have an impact on performance. Table II shows the F1-score under different supports and different tightness. It can be seen that the system performance is slightly better when the smartphone is placed on the table than when it is held in the hand. The reason may be that the smartphone has better stability when placed on the table. But when holding it in the hand, there is a slight difference in the gesture and force of each holding, resulting in a slight reduction in the system performance. For the smartwatch, too-tight strap may make the watch and wrist closely fit together, thus amplifying the impact of wrist on active vibration signal, but the F1-score is still over 90.8%, which indicates that FingerSlid has strong robustness. To improve the performance, we can further collect the data of users in various supports and tightness continuously, and retrain the network parameters regularly.

### D. Performance on User Experience

*1) Sliding Times for Registration:* In the registration phase, the user performs more sliding times can improve the effect of the network training. However, too many times can also lead to an unpleasant user experience. Therefore, we evaluate the FRR of FingerSlid under different sliding times for registration, and the results are shown in Fig. 15. From the figure, we can see that as the number of sliding times increases, the FRR of FingerSlid decreases faster at the beginning. When the user slides 6 and 7 times on the smartphone and smartwatch, the FRR decreases to 1.4% and 1.9%, respectively. Then the FRR continues to decrease at a slow rate until it tends to a stable fluctuation. In order to achieve a balance between system performance and user experience, we fix the sliding times during user registration as 6 on the smartphone, and 7 on the smartwatch in experiments.

*2) Sliding Times for Successful Login:* In the login phase, we evaluate the sliding times required for a legitimate user to successfully login. It can be seen from Fig. 16 that the success rate of one login on the smartphone and smartwatch exceeds 91.7% and 89.6%. Moreover, 96.1% and 93.4% of the login operations on the smartphone and smartwatch can achieve successful authentication within 2 times. Considering the security of the smart device, if 5 consecutive login failures occur, the device is locked for a while.

*3) Response Time for Login:* Since the login includes two stages: the authentication when unlocking the device and the continuous authentication during each interaction, we evaluate the response time of login at different stages as shown in Fig. 17. It can be seen that 95% unlocking authentication and interactive authentication can be completed within 257 ms and 144 ms, re- spectively. The time difference is mainly due to the feature of the loginer needs to be compared with all CBFs in unlocking stage, but at each interaction, the loginer's feature only needs to be compared with one CBF. There is no interaction in the unlocking stage, so the relatively long response time does not affect the user experience. And relevant research [42] shows that the interactive response time within 200 ms can

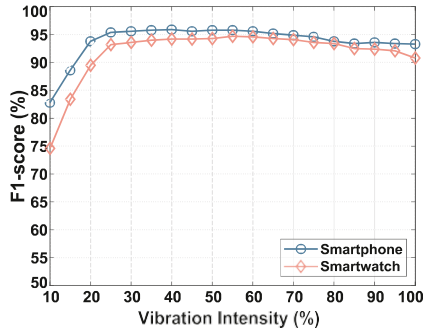

Fig. 18. FRR of sliding times for registration.

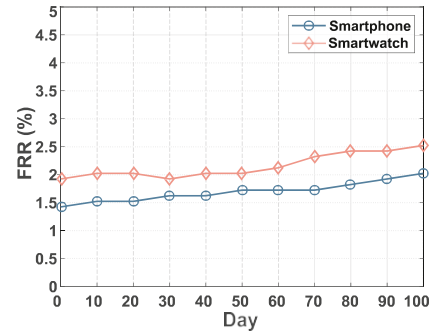

Fig. 20. FRR over different time periods.

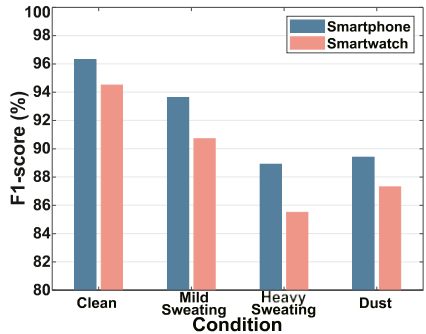

Fig. 19. F1-score under different finger conditions.

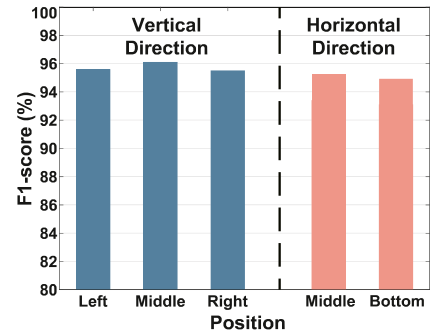

Fig. 21. F1-score under different smartphone positions.

*4) Vibration Intensity:* Since FingerSlid uses active vibration signals to achieve user authentication, we evaluate the effect of vibration intensity on the performance of FingerSlid. We define the maximum vibration intensity of the LRA as 100%. The results under different vibration intensities are shown in Fig. 18. It can be seen that when the intensity reaches 25% and 30%, the F1-score of the smartphone and smartwatch reach the maximum, respectively. Since the vibration signal with a relatively higher intensity can be better received and processed by the accelerometer, and it can resist external noise more effectively. However, when the vibration intensity is too high (exceeds 60%), the influence of fingers on vibration becomes relatively weak, which affects the system performance. To achieve better system performance, we set the vibration intensity of the smartphone to 25% and the smartwatch to 30% in the experiments. And through user research, we find that these two vibration intensities do not affect the daily use of mobile devices.

*5) Finger Conditions:* In order to verify the system performance under different finger conditions, we ask users to authenticate using clean, mildly sweaty, heavily sweaty, and dusty fingers. Fig. 19 shows that compared with clean fingers, heavily sweaty fingers and dusty fingers have a greater impact on system performance. The reason may be that more sweat and dust make the contact surface between the finger and the screen change greatly, thus changing the propagation features of vibration signals. To solve this problem, we can further collect the data of users in various finger conditions and retrain the network parameters to improve the system performance.

*6) Long-Term Performance:* To verify the long-term usability of FingerSlid, we conduct long-term experiments on the smartphone and the smartwatch to evaluate its performance. Our experiments last for 4 months, and the FRR of the data at different times is shown in Fig. 20. The results show that even over time (100 days), the system performance can maintain a very low FRR about 2% and 2.5% on the smartphone and the smartwatch. In order to ensure the longer-term performance of the system, we can continue to periodically train the network model through incremental learning to update the CBFs stored in the system.

### E. Impact of Sliding Behavior

As a behavior-independent authentication system, we need to evaluate the impact of the user's different finger-sliding behaviors on the system, including start-end positions and forces. We use the data under a given behavior as registration data, and the data under other behaviors as test data. There are 5 different start-end positions for users to slide on the smartphone, including left, middle, and right in the vertical direction, and middle and bottom in the horizontal direction. The F1-score of different start-end positions on the smartphone is shown in Fig. 21. It can be seen that when users slide at different start-end positions on the smartphone, the F1-score has no significant difference and are all greater than 95.0%. Due to the small screen of the smartwatch, we only consider the middle positions in horizontal and vertical directions, whose F1-score are 93.3% and 92.8%, respectively.

For the sliding force, users slide on the smartphone and smartwatch with 3 forces (including light, moderate, and hard). Results are shown in Fig. 22, we can see that when the user's sliding force is light or moderate, the highest F1-score on the smartphone and smartwatch can reach 95.1% and 93.6%, respectively. Since the hard sliding force has a greater impact on the vibration signal, the F1-score under hard force on both devices is reduced, but the lowest F1-score also exceeds 91.6%. To further validate the effectiveness of behavior-independent biometric features extracted by FingerSlid, we additionally implement a behavior-dependent system, namely B-FingerSlid, which does not perform behavior-independent feature reconstruction through the Triplet network. The training data of B-FingerSlid is directly from the *Feature Extraction*. We evaluate the F1-score of the two systems under two situations, that is, the registration and login data are from the same behavior, and the registration and login data are from different behaviors (including different start-end positions and forces). The results are shown in Fig. 23. We can see that FingerSlid has high F1-score under all conditions, while B-FingerSlid has a relatively good effect only when the registration and login data come from the same behavior. In particular, the F1-score is reduced to 81.7% when the registration and login data come from different sliding forces.

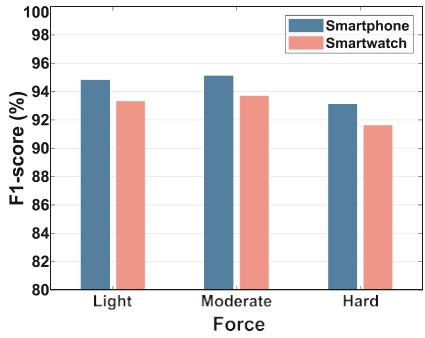

Fig. 22. F1-score under different forces.

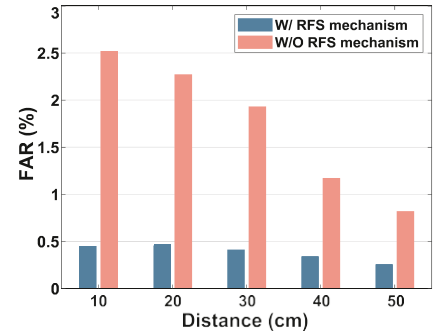

Fig. 24. FAR under replay attacks.

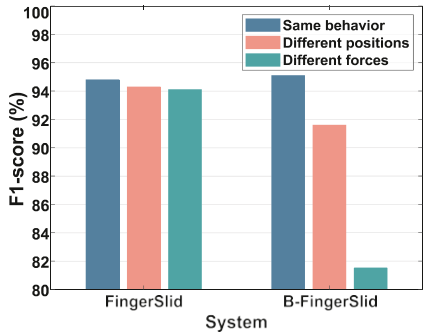

Fig. 23. F1-score under different sliding behaviors.

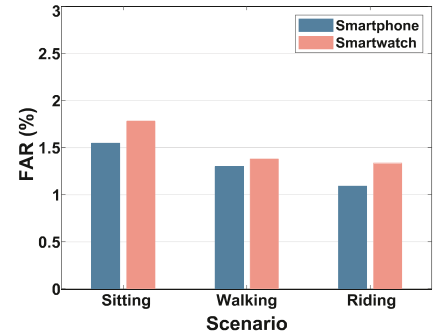

Fig. 25. FAR under mimic attacks.

### F. Performance on Attack Resistance

To evaluate the anti-attack performance of FingerSlid, we conduct experiments under three attacks mentioned in Section III-C.

*1) Resistance to Replay Attack:* Since most of the operations on the smartwatch are performed on the wrist, attackers have little chance to use other devices to record the vibration of the smartwatch. Thus, we only consider replay attacks on the smartphone. Attackers use an extra accelerometer on the same desktop as the legitimate user's smartphone to record the vibration signal, then they use an extra LRA placed on the legitimate user's smartphone to replay the recorded signal for an attack. We conduct experiments by changing the distance between the extra accelerometer and the legitimate user's smartphone. In addition, we also evaluate the resistance to replay attack of our system without the RFS mechanism, and the results are shown in Fig. 24. In the absence of the RFS mechanism in the system, the FAR decreases gradually as the distance between devices increases. When the distance reaches 40 *cm*, the FAR decreases to 1.1%. But no matter at any distance, the FAR of FingerSlid with RFS mechanism does not exceed 0.5%, which indicates that our designed RFS mechanism can effectively resist replay attacks.

*2) Resistance to Mimic Attack:* For the mimic attack, we consider attackers observe and try their best to mimic the sliding behavior of a legitimate user during login in different scenarios. Fig. 25 shows the FAR of mimic attacks on the smartphone and smartwatch under three scenarios. It can be seen that the average FAR of FingerSlid on different mobile smart devices is lower than 1.3%, which indicates that FingerSlid can well resist mimic attacks in various scenarios. The reason is that FingerSlid greatly eliminates the influence of user behavior on authentication by extracting behavior-independent biometric features. Even if attackers can mimic the behavior of a legitimate user, they do not have the same physical features as the legitimate user.

*3) Resistance to Advanced Attack:* Finally, we consider attackers can gain permissions of the LRA and the accelerometer in the legitimate user's device, so they can collect the received signal of each authentication by the accelerometer and filter the RFSs in the received signal. After filtering, the received signal only contains the legitimate user's biometric features. Attackers predict the RFSs of the next sending signal by mathematical modeling (e.g., linear regression), and add predicted RFSs to the received signal to generate new sending signal. Finally, the new signal is sent by the LRA in the legitimate user's device to attack our system. Fig. 26 shows the FAR of the advanced attacks under three scenarios, and the average FAR is about 0.4%, which is similar to the FAR in the mimic attack. Since the RFS added in each sending signal is randomly generated, the information of the RFSs in the next sending signal can not be accurately predicted even by mathematical modeling, so our system can effectively resist advanced attacks.

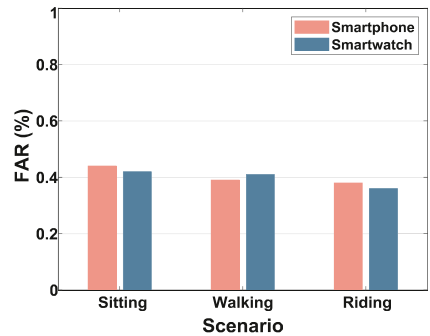

Fig. 26. FAR under advanced attacks.

## VI. DISCUSSION AND LIMITATIONS

We evaluate several factors that affect FingerSlid, but there are also several limitations and opportunities to improve it.

*Security:* FingerSlid mainly collects the unique finger biometrics of users for authentication, which has high security. The RFS mechanism we designed can produce more than two million kinds of RFS combinations, and further prevent potential attack types. However, there are a few uncommon attacks that may pose a threat to system security. For example, the attacker controls the legitimate user's finger to slide while the legitimate user is asleep. In the future, combining with other authentication methods may help to further improve the security of FingerSlid.

*COTS Devices:* Due to the limitations of hardware, we can not control LRAs in COTS mobile smart devices directly for experiments. In order to be as close to COTS devices as possible, we modify COTS devices and use additional LRAs and accelerometers to obtain experimental data. The additional sensor models and locations are consistent with the COTS devices. In the future, if we can obtain the full control of the embedded LRAs in the COTS mobile smart devices, we can deploy FingerSlid on COTS devices without changing the hardware.

*Long-Term Experiments:* Our experiments lasting more than 4 months demonstrate the effectiveness of FingerSlid, but over a longer period, such as several years, the physical features of users' fingers may change, which adversely affect performance of FingerSlid. To solve this problem, we can continue to periodically train the network model through incremental learning to update the physical features stored in the system.

## VII. CONCLUSION

In this article, we propose and implement a continuous authentication system, FingerSlid, for mobile smart devices, which uses active vibration signals to sense the biometric features of users' sliding fingers. We design two different vibration signals, including the chirp signal and the stable signal, to extract different biometric features, and then design a Triplet network to reconstruct the behavior-independent biometric features. Finally, we use the CBF to achieve user authentication. In addition, we design a novel RFS mechanism to improve the anti-attack ability of FingerSlid. A large number of experiments on the smartphone and smartwatch under various conditions show that FingerSlid achieves an average authentication accuracy of 95.4%, and 99.5% of attacks can be resisted.

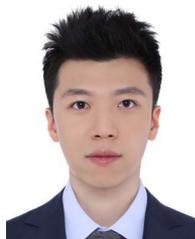

**Yadong Xie** (Graduate Student Member, IEEE) received the BE degree in network engineering from Hebei University, China, in 2016. He is currently working toward the PhD degree with the School of Computer Science, Beijing Institute of Technology, Beijing, China. His research interests include mobile computing, mobile health, human-computer interaction, and deep learning.

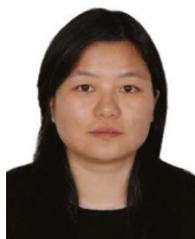

**Fan Li** (Member, IEEE) received the BEng and MEng degrees in communications and information system from the Huazhong University of Science and Technology, China, in 1998 and 2001, respectively, the MEng degree in electrical engineering from the University of Delaware, in 2004, and the PhD degree in computer science from the University of North Carolina at Charlotte, in 2008. She is currently a professor with the School of Computer Science, Beijing Institute of Technology, China. Her current research focuses on wireless networks, ad hoc and sensor networks, and mobile computing. Her papers won Best Paper Awards from IEEE MASS (2013), IEEE IPCCC (2013), ACM MobiHoc (2014), and Tsinghua Science and Technology (2015). She is a member of ACM.

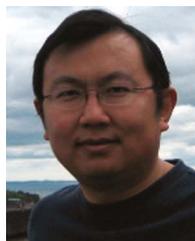

**Yu Wang** (Fellow, IEEE) received the BEng and MEng degrees in computer science from Tsinghua University, and the PhD degree in computer science from the Illinois Institute of Technology. He is currently a professor with the Department of Computer and Information Sciences, Temple University. His research interest includes wireless networks, smart sensing, and mobile computing. He has published more than 200 papers in peer reviewed journals and conferences, with four best paper awards. He has served as general chair, program chair, program committee member, etc. for many international conferences (such as IEEE IPCCC, ACM MobiHoc, IEEE INFOCOM, IEEE GLOBECOM, IEEE ICC). He has served as editorial board member of several international journals, including *IEEE Transactions on Parallel and Distributed Systems*. He is a recipient of Ralph E. Powe Junior Faculty Enhancement Awards from Oak Ridge Associated Universities (2006), Outstanding Faculty Research Award from College of Computing and Informatics at the University of North Carolina at Charlotte (2008) and ACM distinguished member (2020).